\documentclass[a4paper,10pt]{article}

\title{Some Aspects of Virtual Black Holes}
\author{Mir Faizal\\Department of Mathematics,
University of Durham\\
 DH1 3LE, Durham,  United Kingdom \\faizal.mir@durham.ac.uk}

\begin{document}

\maketitle

\begin{abstract} In this paper we shall consistently third quantize modified gravity. 
 Then we shall analyse certain aspects of virtual black holes in this third quantized modified gravity. 
We will see how 
 a statistical mechanical origin for  the  Bekenstein-Hawking entropy naturally arises in this model.
 Furthermore, in this model
the area and thus the entropy of a real macroscopic black hole is quantized.
Virtual black holes 
cause loss of quantum coherence and this gives an intrinsic entropy to all physical systems which can be used to 
define a direction of time and hence provide
a solution to the problem of time.
\end{abstract}



\section{Introduction:-} It is expected that the quantum fluctuations of spacetime  can cause the topology  of spacetime to  change  at the 
Planck scale and thus giving it a foam like structure called the spacetime foam \cite{2, 3}. Spacetime form has largely been discussed 
via the formation of  baby universes  which render the spacetime 
 multiple connected \cite{4, 4a, 4b}.  So in this model the spacetime manifold has a large value of first Betti  
number $B_1$ and  the second Betti number vanishes,
 $B_2 =0$. 
The problem with this model is that it predicts  a  wrong value of Q.C.D. $ \theta$-parameter \cite{6} and the
 cosmological constant \cite{6a, 6b}. 

However, there is an alternative 
model of spacetime foam which seems to predict a correct value of the Q.C.D. $\theta$-parameter \cite{5}. 
In  this model the topology of spacetime changes by the formation of virtual black holes and thus the spacetime 
remains single connected \cite{5a, 5b}. 
So in this model the spacetime manifold has a large 
value of second  Betti  number $B_2$ and the first and third Betti numbers vanish,
  $B_1 = B_3 = 0$.  In 
this picture there is also an elegant way to describe black hole evaporation without  the appearance of a naked singularity.
 Macroscopic real black holes evaporate down to Planck size by radiating
 Hawking radiation. At this stage they are left with no energy or charge. They then disappear in a sea of virtual black holes. 
As this picture seems 
to be a more realistic picture of spacetime foam, we will analyse certain aspects of it in this paper. 

To study 
the physical effects of virtual black holes 
one should have analysed the collision of particles
with energy less than the 
Planck energy in a small region  containing a virtual
black hole.  To do so we would need to  find a Euclidean solution for  
this process.  However, it is very difficult to find such a solution. 
So,  we shall analyses virtual black holes 
via third quantization.  
The third quantization  has been 
 discussed implicitly in Refs. \cite{2b,3b} 
 and explicitly in Refs.  \cite{2ba,3ba}. The modification of Wheeler-DeWitt equation 
by the addition of 
non-linear terms and the third quantization of the resultant theory was formally analyzed  in Ref. \cite{sg}. Third quantization of  
 Brans-Dicke 
theories  \cite{i} and Kaluza-Klein theories  \cite{ia} has also been done.  
However, all this work has been done in the baby universe model 
of spacetime foam.   
 We shall therefore apply third quantization 
to virtual black hole model in this paper.
It may be noted that the idea canonical 
quantization of gravity  has progressed  into
loop
quantum gravity \cite{lqg, aalqg}. 
Furthermore, the idea of third quantization now appears as  
group field theory \cite{gft, gft1} in loop quantum gravity. 
Hense, this present work should be translated 
into the language of 
group field theory. However, 
  it is not clear how to  deal with virtual black holes in 
group field theory. To understand
 that 
it might be useful to first analyse  virtual black 
holes in two dimensions via matrix models \cite{mm1, mm2}. 
This is because group field theory can be 
viewed as a higher dimensional generalisation of
 the matrix models.    
 
\section{Wheeler-DeWitt Equation}
It is hoped that a corrected gravitational potential could
fit galaxy rotation curves without the need of dark matter \cite{g1, g2}. So $f(R)$ gravity theories have become very important \cite{g3, g4}. 
We will thus study the virtual black holes in $f(R)$ gravity theories. 
The Lagrangian density describing a generic $f(R)$ theory
of gravity is given by
\begin{eqnarray}
\mathcal{L}&=&\sqrt{-g}\left(  f\left(  R\right)  -2\Lambda_{c}\right),
\nonumber  \\ f^{\prime\prime}&\neq& 0,\label{lag}
\end{eqnarray}
where $f\left(  R\right)  $ is an arbitrary  function of the scalar
curvature and primes denote differentiation with respect to the scalar
curvature.  The Hamiltonian constraint for $f(R)$ gravity is given by~\cite{cr}
\begin{eqnarray}
\mathcal{H}&=&\frac{1}{2\kappa}\left[  \frac{\mathcal{P}}{6}\left(  {}^{\left(
3\right)  }R-2\Lambda_{c}-3K_{ij}K^{ij}+K^{2}\right) \right. \nonumber \\ && \left. +V(\mathcal{P})-\frac
{1}{3}g^{ij}\mathcal{P}_{\mid ij}-2p^{ij}K_{ij}\right],
\end{eqnarray}
where
\begin{equation}
V(\mathcal{P})=\sqrt{h}\left[  Rf^{\prime}\left(  R\right)  -f\left(
R\right)  \right]  .\label{V(P)},
\end{equation}
and 
\begin{equation}
 G_{ijkl} = \frac{1}{2 \sqrt{h}}(h_{ik}h_{jl} +h_{il}h_{jk} - h_{ij}h_{kl}  ).
\end{equation}
Here $K_{ij}$ is the second fundamental form and $K = h^{ij}K_{ij}$ is its trace 
and ${^{(3)}}R$ is the three dimensional scalar curvature. 
As we have 
\begin{equation}
\mathcal{P=}-6\sqrt{h}f^{\prime}\left(  R\right).
\end{equation}
So, we get
\begin{eqnarray}
\mathcal{H}&=&\frac{1}{2\kappa}\left[  -\sqrt{h}f^{\prime}\left(  R\right)
\left(  {}^{\left(  3\right)  }R-2\Lambda_{c}-3K_{ij}K^{ij}+K^{2}\right)
+V(\mathcal{P})\right.\nonumber \\&&\left. +2g^{ij}\left(  \sqrt{h}f^{\prime}\left(  R\right)  \right)
_{\mid ij}-2p^{ij}K_{ij}\right]  .\label{Hamf(R)}
\end{eqnarray}
This can be expressed as  
\begin{eqnarray}
\mathcal{H}&=& f^{\prime}\left(  R\right)  \left[  \left(  2\kappa\right)
G_{ijkl}\pi^{ij}\pi^{kl}{}-\frac{\sqrt{h}}{2\kappa}\left(  ^{\left(
3\right)  }R -2\Lambda_{c}\right)  \right]\nonumber \\  && +\frac{1}{2\kappa}\left[  \sqrt
{g}f^{\prime}\left(  R\right)  \left(  2K_{ij}K^{ij}\right)  +V(\mathcal{P}%
)  +2g^{ij}\left(  \sqrt{h}f^{\prime}\left(  R\right)  \right)_{\mid
ij}\right.\nonumber \\&&\left.  -2p^{ij}K_{ij}\right]  .
\end{eqnarray}
Now by using 
\begin{equation}
p^{ij}=\sqrt{h}K^{ij},
\end{equation}
and transforming into canonical momenta the Hamiltonian constraint for $f(R)$ gravity becomes
\begin{eqnarray}
\mathcal{H}&=&f^{\prime}\left(  R\right)  \left[  \left(  2\kappa\right)
G_{ijkl}\pi^{ij}\pi^{kl}{}-\frac{\sqrt{h}}{2\kappa}{}\left(  ^{\left(
3\right)  }R-2\Lambda_{c}\right)  \right]\nonumber \\&& +\left(  4\kappa\right)  \left[
G_{ijkl}\pi^{ij}\pi^{kl}+\frac{\pi}{4}^{2}\right]  \left(  f^{\prime}\left(
R\right)  -1\right) \nonumber \\&&  +\frac{1}{2\kappa}\left[  V(\mathcal{P})+2g^{ij}\left(
\sqrt{h}f^{\prime}\left(  R\right)  \right)  _{\mid ij}\right]
.\label{Hamf(R)_1}
\end{eqnarray}
Wheeler-DeWitt equation is the quantum mechanical version of this Hamiltonian constraint 
\begin{equation}
 H \phi(h) =0,\label{sysysy}
\end{equation}
where we have used 
\begin{equation}
  \pi^{ij}  = - i \frac{\delta}{\delta h_{ij}}.
\end{equation}
It may be noted that when  $f\left(  R\right)  =R$, then  $V(\mathcal{P}
)=0$ and the Wheeler-DeWitt equation for $f(R)$ gravity reduces to 
the usual Wheeler-DeWitt equation.
In most interpretations of quantum gravity e.g., naive \cite{18}, conditional probability \cite{19}, WKB 
approximation \cite{20}, Wheeler-DeWitt equation is analogous to  Schroedinger wave equation,
 in the sense it represents the quantum state of a 
single universe.  

However, in the  third quantized formalism it is seem as a classical field equation that has to be third quantized \cite{2ba,3ba}. 
So  third quantized formalism  describes the quantum state of an ensemble of geometries. Thus third quantized formalism is the natural  
formalism 
for analysing any model of the  spacetime foam. Lot of work on third quantization is done  in
analogy with  quantum field theory in flat spacetime \cite{2b,3b, sg}. 
 However,  there is no timelike Killing vector for  Wheeler-DeWitt equation \cite{til, til1}.
 So consistent third quantization should be done in analogy with quantum field
 theory in curved spacetime and this will be done in the next section. 

 \section{Third Quantization}
In this section we will third quantize the Wheeler-DeWitt equation for $f(R)$ gravity. So we first interpret Eq. $(\ref{sysysy})$ as a
 classical field 
equation and then quantize it. To do so,
let us assume that $\{\phi(P, h) \}$ and $\{\phi^*(P, h) \}$ form a complete set of solutions to this Wheeler-DeWitt equation, and  satisfies  
a Klein-Gordon type  symplectic product with the following properties, 
\begin{eqnarray}
 \int Dh \mathcal{J}(\phi(P, h) ,\phi(Q, h) ) &=& \mathcal{M}(P,Q) , \label{conditions1}\\ 
\int Dh \mathcal{J}(\phi(P, h) ,\phi^*(Q, h) ) &=& 0, \label{conditions2}\\ 
\int Dh \mathcal{J}(\phi^*(P, h) ,\phi^*(Q, h) ) &=& - \mathcal{M}(P,Q) .\label{conditions3}
\end{eqnarray}
In quantum field theory the   condition given in Eq. $(\ref{conditions2})$ will not hold in general and so this is a requirement on the 
complete set of solutions to the Wheeler-DeWitt equation \cite{bd}. 
We also chose  $\mathcal{M}(P,Q) $ to have positive eigenvalues only. This again is not always
 true and so this is again a requirement on the complete set of solutions to the Wheeler-DeWitt equation.

In third quantized formalism  $\phi(h)$ is promoted to an Hermitian operators and expressed as \cite{sg}
 \begin{equation}
  \hat{\phi}(h) = \int DP  [a(P)  \phi(P, h) + a^{\dagger}(P)  \phi^*(P, h) ], \label{bov}
 \end{equation}
where $a(P)$ and $a^{\dagger}(P)$ satisfy 
\begin{eqnarray}
 \left[a(P), a^{\dagger}(Q)   \right] &=& \delta (P, Q), \nonumber \\ 
 \left[a^{\dagger}(P), a^{\dagger}(Q)   \right] &=& 0,  \nonumber \\ 
 \left[a(P), a(Q)   \right] &=& 0,
\end{eqnarray}
where $\delta(P, Q)$ is defined by 
\begin{equation}
 \int DP \delta(P, Q) \phi(P, h) = \phi(Q, h). 
\end{equation}
For this choice of complete set of solutions to the  Wheeler-DeWitt equation, 
Eq.~$(\ref{sysysy})$,
 we define a state called the vacuum state $|0\rangle$, as the state  that  is annihilated by $a(P) $:
 \begin{equation}
  a(P)  |0\rangle = 0.
 \end{equation}
Now $a^{\dagger}(P) $ and $a(P) $ can be called the creation and annihilation operators respectively, 
in analogy  with those for the simple quantum harmonic oscillator. 
They will create and annihilate geometries in the third quantized formalism. 

It may be noted that  the division between $\{\phi(P, h) \}$ and $\{\phi^*(P, h) \}$ is not unique even after imposing conditions given by
 Eqs. $(\ref{conditions1})$-$(\ref{conditions3})$  $\cite{bd}$. Due to this non-uniqueness in division between
 $\{\phi(P, h) \}$ and $\{\phi^*(P, h) \}$, there is   non-uniqueness in the definition of the vacuum state also. 
This can be seen by considering  $\{\phi'(P, h) \}$ and $\{\phi'^*(P, h) \}$ as another  
 complete set of solutions to  Eq. $(\ref{sysysy})$, satisfying
 conditions given by Eqs. $(\ref{conditions1})$-$(\ref{conditions3})$. Now we have 
\begin{equation}
   \hat{\phi}(h) = \int DP  [a'(P)  \phi'(P, h) + a'^{\dagger}(P)  \phi'^*(P, h)]. \label{bov1}
\end{equation}
Here the  vacuum state $|0'\rangle$ is   the state annihilated by $a'(P) $,
\begin{equation}
 a'(P) |0'\rangle =0.
\end{equation}
Now many geometry states can be built  by repeated action of $a'^{\dagger}(P) $ on $|0'\rangle$.
As $\phi(P, h) $ and $\phi^*(P, h) $ form a complete set of  solutions to the field equation, Eq.~$(\ref{sysysy})$, we can 
 express $\phi'(P, h) $  
as a linear combination of $\phi(P, h) $ and $\phi^*(P, h) $, 
\begin{equation}
 \phi'(P, h) = \int DQ  [\alpha(P,Q) \phi(Q, h) + \beta(P,Q) \phi^*(Q, h)].\label{bovv2} 
\end{equation}
By substituting Eq. $(\ref{bovv2})$ in Eq. $(\ref{bov1})$ and comparing the resulting expression with Eq. $(\ref{bov})$, we find 
\begin{eqnarray}
 a(P)  &=& \int DQ  [\alpha(P,Q) a'(Q)  + \beta^*(P,Q) a'^{\dagger}(Q) ],\\
a^{\dagger}(P)  &=& \int DQ  [\alpha^*(P,Q)  a'^{\dagger}(Q)  + \beta(P,Q) a'(Q) ].
\end{eqnarray}
The two Fock spaces based on these choices of complete set of  solutions to the field equation, 
Eq. $(\ref{sysysy})$, are different as long as $\beta(P,Q)  \neq 0$. In particular $a(P) |0'\rangle$ does not vanish because  
\begin{eqnarray}
  a(P) |0'\rangle &=& \int DQ  [\alpha(P,Q) a'(Q)  + \beta^*(P,Q) a'^{\dagger}(Q) ]|0'\rangle  \nonumber \\
&=&\int DQ  \beta^*(P,Q) a'^{\dagger}(Q)  |0'\rangle \neq 0,
\end{eqnarray}
but, 
\begin{equation}
 a(P)  |0\rangle =0.
\end{equation}
Thus $a(P) |0'\rangle$ is  a one-geometry state. In fact we have 
\begin{equation}
\langle 0'| a(P) ^{\dagger} a(P) |0'\rangle = \int DU DQ \beta(P,U) 
\beta^*(P,Q) \mathcal{M} (U,Q).
\end{equation}
The Wightman two-point function  is now given by
\begin{equation}
 G(h, h') = \langle 0| \hat{\phi}(h)  \hat{\phi}(h')|0\rangle.
\end{equation}
This can be written as:
\begin{eqnarray}
 G(h, h') &=& \int D P D  Q \langle 0|(a(P)  \phi(P, h) +  a(P) ^{\dagger}\phi^*(P, h))\nonumber \\ && 
\times  (a(Q) \phi(Q, h') + a^{\dagger} (Q)  
\phi(Q, h')|0\rangle  \nonumber \\ &=&\int D P D Q \phi(P,h) \phi^*(Q, h')\mathcal{C}(P,Q) ,
\end{eqnarray}
where $\mathcal{C}(P,Q) $ is the commutator,
\begin{equation}
 \mathcal{C}(P,Q)  = \langle 0|[a(P) , a^{\dagger}(Q) ]|0\rangle. \label{tion}
\end{equation}
Now as $\hat{\phi}$ is Hermitian, so we have \cite{bd}
 \begin{equation}
[(\phi(P) , \hat{\phi}),(\hat{\phi}, \phi(Q) )]   = \mathcal{M}(P,Q) ,\label{bov7}
\end{equation}
 and so 
\begin{eqnarray}
  (\phi(P) , \hat{\phi}) &=&  \int D U  a(U)  \mathcal{M}(P,U) ,  \label{bov2} \\ 
 ( \hat{\phi}, \phi(Q) ) &=&  [(\phi(Q) , \hat{\phi})]^{\dagger}  =   
 \int D W  a^{\dagger}(W)  \mathcal{M}(W,Q). \label{bov6}
\end{eqnarray}
Now from Eqs. $(\ref{bov7})$-$(\ref{bov6})$, we get 
\begin{equation}
 \int D U D W \mathcal{M}(P,U) [a(U) , a^{\dagger}(W) ]\mathcal{M}(W,Q) = \mathcal{M}(P,Q) .\label{tion1}
\end{equation}
Using Eqs. $(\ref{tion})$ and $(\ref{tion1})$, we get
\begin{equation}
  \int D U D W \mathcal{M}(P,U) \mathcal{C}(U,W)\mathcal{M}(W,Q) = \mathcal{M}(P,Q) .
\end{equation}
 This equation  in matrix notation is written as, 
\begin{equation}
 \mathcal{M}\mathcal{C}\mathcal{M} = \mathcal{M}.
\end{equation}
Assuming that $\mathcal{M}(P,Q) $ has only positive eigenvalues, i.e., that it is  invertible, we get
\begin{equation}
 \mathcal{C}= \mathcal{M}^{-1}.
\end{equation}
 Therefore the two-point function is given by
 \begin{equation}
   G(h,h') = \int DP DQ  \phi(P, h) \phi^*(Q, h') \mathcal{M}^{-1}(P,Q). \label{twopointfunctionscalar}
 \end{equation}
In this section we developed a third quantization of the Wheeler-DeWitt equation for $f(R)$ gravity. 
In the next section we will use it to analyse the formation of virtual black holes. 
\section{Virtual Black Holes}
Wheeler-DeWitt equation in third quantized 
formalism represents 
the quantum state of an ensemble of non-interacting geometries. 
However, this is still not enough to account for topology change.
 To obtain a theory consistent with topology change we need to
 include interaction terms.
So we modify the original Wheeler-DeWitt equation by the addition of interaction terms, 
\begin{equation}
 \left[  H\phi  - \frac{\delta \mathcal{V} [\phi]}{\delta \phi (h)}\right] =0,
\end{equation}
where  $\mathcal{V}[\phi]$ is the potential that summarizing all the interactions. 
We can now apply this third quantized formalism of quantum gravity to  
virtual black holes. 
It has been argued that virtual black holes may form in loops like 
other virtual particles form in the conventional quantum field 
theory \cite{5}.
However, this discussion on  virtual black hole loops so far has not been made precise. 
This can be  very easily done in the third 
quantized formalism. 
We can thus write the amplitude for
 the formation of a virtual black hole loop in third quantized formalism as: 
\begin{equation}
  \mathcal{A} = \int Dh Dh'  G(h,h')G(h',h).\label{twopointfunction1scalar1}
\end{equation}
 Now from Eqs. $(\ref{twopointfunctionscalar})$ and $(\ref{twopointfunction1scalar1})$, we get   
\begin{eqnarray}
   \mathcal{A} &=&   \int Dh Dh' DP DQ DU DW \nonumber \\ && \times \phi(P, h) \phi^*(Q, h') \mathcal{M}^{-1}(P,Q) \nonumber \\ && \times 
\phi(U, h') \phi^*(W, h) \mathcal{M}^{-1}(U,W).
\end{eqnarray}
Thus these virtual black holes form as off-shell contributions in the third quantized formalism of quantum gravity.  

Virtual black holes will exist in dynamic equilibrium with 
flat spacetime at Planck scale. However,
 if due to any process 
$\mathcal{A}^2$ i.e., the probability of formation of virtual black holes  increases then 
 at a certain critical value of $\mathcal{A}^2$ a phase transition can occur. This can  
change the large scale structure of spacetime and create a real black hole. Thus the formation of a real black hole can be seen as 
 phase transition induced 
by virtual black holes in analogy to how a ferromagnet gets magnetised. This implies  that the spacetime foam can generate the micro-sates 
of a real 
black hole. 
  
So we assume that the macroscopic
 state of a black hole is made up of micro-states of spacetime foam. 
Thus the area of a real macroscopic black hole  $A$ is covered by micro-states of foam like structure of spacetime at the Planck's scale.
So  if $N$ is the number of 
micro-states  that completely cover the 
area of this macroscopic black hole at the Planck scale then  the total statistical mechanical entropy associated with the real black hole
 because
 of these micro-states is 
proportional to $N$ \cite{sme}. This suggest the fact that the entropy of a real black hole can have a statistical mechanical origin and the 
foam like structure of spacetime may give rise to the micro-states for the Bekenstein-Hawking entropy. 

In fact as $A$ is completely covered by these micro-states of spacetime foam, we have 
\begin{equation}
 A = N \tilde{A} = 4 N \pi,  
\end{equation}
 where $\tilde{A} = 4\pi$ is a unite Planckian area   in the Planck's unites. 
As the macroscopic black hole area is represented by $N$ microscopic areas of  Planckian size it is naturally quantized. 
Now the Bekenstein-Hawking entropy along with the quantum corrections to it for a macroscopic black hole is given by \cite{bhe}
\begin{eqnarray}
 S &=& \frac{A}{4} + \alpha_0 \log A +  \frac{\alpha_1}{A} + \frac{\alpha_2}{A^2} + \cdots  \nonumber \\ &=& 
N \pi  + \alpha_0 \log  (4 N \pi) +  \frac{\alpha_1}{4 N \pi} + \frac{\alpha_2}{16 N^2 \pi^2} + \cdots,
\end{eqnarray}
where $\alpha_0, \alpha_1, \alpha_2 \cdots$ are constants.  
Now as $N$ is a very large number we can neglect the higher order corrections to this Bekenstein-Hawking entropy.
Thus to the leading order this entropy is   
\begin{equation}
 S\sim N\pi,
\end{equation}
 and so to the leading order the Bekenstein-Hawking entropy is proportional to $N$.
 This means that the entropy of a real black hole is  quantized by the  structure of spacetime foam. 

\section{Quantum Coherence and the Problem of Time } 
Now  there will always be some probability for the formation of virtual black holes in flat spacetime. So the particles found in nature 
will naturally
interact with  virtual black holes even in flat spacetime. 
Now if $\mathcal{H}_{phy}$ is the Hilbert space of the particles found in nature and $\mathcal{H}_{vb}$ is the Hilbert space of these 
virtual black holes 
then the total Hilbert space $\mathcal{H}$ for this physical theory  will be given by 
\begin{equation}
 \mathcal{H} = \mathcal{H}_{phy}\otimes  \mathcal{H}_{vb}. 
\end{equation}
Density operator $\rho$ can now be expressed as   $|\Psi\rangle\langle\Psi|$
where $|\Psi\rangle$ is a vector or total wave function in $\mathcal{H}$. It is a pure  state.

Now if $\rho_-$ and $\rho_+$ are density matrices for the total Hilbert space  at past and future infinity respectively and ${\$}$
is the superscattering operator, than we can write \cite{ss}
\begin{equation}
\rho_+ = {\$}\rho_-.
\end{equation}
Now as the set of creation operators for the total Hilbert space $\mathcal{H}$ form a complete set of bases at both past and future 
infinity, we can write ${\$}$ as 
\begin{equation} 
{\$} = \mathcal{S}\mathcal{S}^{\dagger},
\end{equation}
where  $\mathcal{S}^{\dagger}$ is the adjoint of the $\mathcal{S}$-matrix. 
Now to see how the density matrix evolves we have to take the trace of the future density matrix, which is given by
\begin{equation}
Tr(\rho_+^2)=Tr(({\$}\rho_-)({\$}\rho_-)).
\end{equation}
As the supperscattering matrix factorises so we can write
\begin{equation}
Tr(({\$}\rho_-)({\$}\rho_-)) = Tr(\mathcal{S}\rho_-\mathcal{S}^{\dagger}\mathcal{S}\rho_-\mathcal{S}^{\dagger}).
\end{equation}
Thus we have
\begin{equation}
Tr(\mathcal{S}\rho_-\mathcal{S}^{\dagger}\mathcal{S}\rho_-\mathcal{S}^{\dagger}) 
= Tr(\mathcal{S}\rho^2_-\mathcal{S}^{\dagger}) = Tr(\rho_-^2).
\end{equation}
This is the trace of the past density matrix. 

However, the states of virtual black holes are not measurable so in reality we have to take the partial trace over $\mathcal{H}_{vb}$.
Now if $I_{vb}$ is the identity operator on $\mathcal{H}_{vb}$, and $A_{phy}$ is an observable in $\mathcal{H}_{phy}$,
 then physically meaningful measurements will be given by 
\begin{equation}
 Tr (\rho I_{vb}\otimes A_{phy}).
\end{equation}
As  the states for $\mathcal{H}_{phy}$ do not form a complete set of bases by themselves and so the superscattering 
operator does not factorise into $\mathcal{S}$-matrix 
and its adjoint thus the evolution for $\mathcal{H}_{phy}$ is non-unitary. This  causes loss of quantum coherence. 

The total wave function $|\Psi\rangle$ may be written as a superposition, with
coefficients $c_n $ satisfying
\begin{equation}
 \sum_n  |c_n |^2=1,
\end{equation}
 of single tensor
products  
\begin{equation}
 |\Psi\rangle=\sum_n  c_n |vb_n \rangle\otimes|phy_n \rangle,
\end{equation}
where the $|vb_n \rangle$ and $|phy_n \rangle$ are orthonormal sets of basis vectors in $\mathcal{H}_{vb}$ and $\mathcal{H}_{phy}$
respectively.  Thus we get
\begin{eqnarray}
\rho_{vb}&=&\sum_n 
|c_n |^2|vb_n \rangle\langle vb_n |, \nonumber \\ 
\rho_{phy}&=&\sum_n  |c_n |^2 |phy_n \rangle\langle phy_n |. 
\end{eqnarray}
The von Neumann for the physical system is given by 
\begin{equation}
 S_{phy} = -Tr(\rho_{phy}\log \rho_{phy}).
\end{equation}
Thus we can write 
\begin{equation}
 S_{phy} = -\sum_n |c_n |^2\log|c_n |^2.
\end{equation}
This is the entropy that will be associated with the physical systems. 

Now this entropy can be used to define a direction of time.
 In order to proceed to define time like  
 anything on the  traditional lines, one
will need a notion for a flow of time, represented by a one-parameter
family of  unitary operators, which we shall call a  flow
$t\mapsto U(t)$ on $\mathcal{H}$ with $t$ ranging over the non-negative real
numbers, mapping any initial density matrix 
$\rho_0=|\Psi_0\rangle\langle\Psi_0|$ at some initial time $t=0$ into
the density matrix  $\rho_t=|\Psi_t\rangle\langle\Psi_t|$ at a later time,
$t$, according to the transformation 
\begin{equation}
 \rho_t=U(t)\rho_0 U(t)^{-1}.
\end{equation}
So a single microstate at one time evolves to a single  microstate at a
later time. Now for any initial state the entanglement between $\mathcal{H}_{vb}$ and 
$\mathcal{H}_{phy}$ will be less than what it will be at a later stage. As the states will keep getting more and more entangled with 
the passage of time,
so this entanglement can also be used to identify the direction of time. Thus the value of entropy will also 
increase uniformly as a state evolves in future. Thus we can equate the direction of increase of this von Neumann entropy $S(t)_{phy}$ with 
time. 
This can thus give a solution to the problem of time in quantum gravity. 

\section{Conclusion}
In this paper  modified gravity is consistently third quantized  in analogy with the quantization of  scalar field theory in curved spacetime. 
Then the virtual black holes model of spacetime foam, which at present seems to be the  correct model of spacetime foam 
is analysed in this third quantized modified gravity.   
This model is used to give a statistical origin of the  Bekenstein-Hawking entropy. It is also shown that the 
  the area
and thus the entropy of a real black hole  is quantized in this model.  
Furthermore, the loss of quantum coherence  occurs  due to the fact that virtual black hole states are not measurable.  This in turn causes 
 all physical systems  
to get an intrinsic entropy. This entropy is used to give a definition of time in 
quantum gravity.

It will be interesting to  analysis  many other results that have been discussed 
  for baby universes in the third quantized formalism to spacetime foam formed by virtual
 black holes. It might be 
possible to get a different value of the cosmological constant here than what was obtained in the model of spacetime foam containing baby 
universes. 
In higher dimensions  spacetime is known to possess more exotic topologies like the black rings. It will also be interesting to analyse a 
model of spacetime containing  virtual black rings. Results of this paper can be easily be generalised  to virtual black rings.

\end{document}